\documentclass[12pt,a4paper]{article}
\usepackage{amssymb,amsmath}
\usepackage{cite}
\usepackage[dvips]{lscape,graphicx}

\DeclareMathOperator{\rre}{Re} \DeclareMathOperator{\iim}{Im}

\begin{document}

\author{R. B. Nevzorov${}^{\dag,\ddag}$, K. A. Ter--Martirosyan${}^{\dag}$, and M. A. Trusov${}^{\dag}$ \\[5mm]
{\itshape ${}^{\dag}$ITEP, Moscow, Russia} \\ {\itshape
${}^{\ddag}$DESY Theory, Hamburg, Germany}}

\title{Higgs bosons in the simplest SUSY models }

\maketitle

\begin{abstract}
\noindent Nowadays in the MSSM the moderate values of $\tan\beta$
are almost excluded by LEP\,II lower bound on the lightest Higgs
boson mass. In the Next--to--Minimal Supersymmetric Standard Model
the theoretical upper bound on it increases and reaches maximal
value in the strong Yukawa coupling limit when all solutions of
renormalization group equations are concentrated near the
quasi--fixed point. For calculation of Higgs boson spectrum the
perturbation theory method can be applied. We investigate the
particle spectrum in the framework of the modified NMSSM which
leads to the self-consistent solution in the strong Yukawa
coupling limit. This model allows one to get $m_h\sim
125\text{~GeV}$ at values of $\tan\beta\ge 1.9$. In the
investigated model the lightest Higgs boson mass does not exceed
$130.5\pm 3.5\text{~GeV}$. The upper bound on the lightest
CP--even Higgs boson mass in more complicated supersymmetric
models is also discussed.
\end{abstract}

\newpage

\section{Introduction}

Last year there was a great progress in the Higgs boson searches.
The experimental lower bound on the Higgs boson mass in the
Standard Model (SM) has increased from $95.2\text{~GeV}$ \cite{1}
to $113.3\text{~GeV}$ \cite{2}. At the same time the upper bound
that comes from an analysis of radiative corrections to the
electroweak observables has diminished to $210\text{~GeV}$
\cite{2}. Thus the allowed region of the Higgs boson mass in the
SM has shrunk drastically. Moreover at the LEP\,II a few
additional $b\bar{b}$ events were observed \cite{3}. They can be
interpreted as a signal of the Higgs boson production with mass
$115\text{~GeV}$ in the $e^{+}e^{-}$ annihilation. Such Higgs
boson mass does not agree with theoretical lower bound in the SM
following from the stability of the physical vacuum up to the
Planck scale $M_{\text{Pl}}\approx 2.4\times 10^{18}\text{~GeV}$
[\citen{4}--\citen{7}].

The simplest supersymmetric (SUSY) extension of SM is the Minimal
Supersymmetric Standard Model (MSSM). Its Higgs sector includes
two doublets $H_1$ and $H_2$. Each of them after spontaneous
symmetry breaking acquires a nonzero vacuum expectation value
$v_1$ and $v_2$, respectively. Instead of them, the sum of their
squares $v^2=v_1^2+v_2^2$ and the value of $\tan\beta=v_2/v_1$ are
usually used.

An important feature of supersymmetric models is the existence of a light Higgs boson
in the CP--even Higgs sector. The upper bound on its mass strongly depends on the value
of $\tan\beta$. At the tree level the lightest Higgs boson mass does not exceed \cite{8} the
$Z$--boson mass: $m_h\le M_Z|\cos 2\beta|$. The loop corrections from the
$t$--quark and its superpartners significantly raise the upper bound on $m_h$:
\begin{equation}
\label{1} m_h\le\sqrt{M_Z^2\cos^2 2\beta+\Delta}, \end{equation}
where $\Delta$ in the one--loop approximation is given by
\begin{equation}
\label{2}
\Delta\approx\frac{3}{2\pi^2}\frac{m_t^4}{v^2}\left[\ln\frac{M_S^2}{m_t^2}+
\frac{X_t^2}{M_S^2}\left(1-\frac{X_t^2}{12 M_S^2}\right)\right].
\end{equation}
Here $m_t$ is the running top quark mass at the electroweak scale
(at $q=M_t^{\text{pole}}=174\text{~GeV}$, $M_t^{\text{pole}}$ is
the $t$--quark pole mass), $X_t$ is the stop mixing parameter, and
$M_S$ is the SUSY breaking scale which is expressed via the stop
masses $m_{\tilde{t}_1}$ and $m_{\tilde{t}_2}$:
$M_S=\sqrt{m_{\tilde{t}_1}m_{\tilde{t}_2}}$. The one--loop
corrections (\ref{2}) attain maximal value at
$X_t=\pm\sqrt{6}M_S$. These corrections are proportional to
$m_t^4$ and depend logarithmically on the SUSY breaking scale
$M_S$. They are almost insensitive to the choice of $\tan\beta$.
The absolute value of $\Delta$ is of order of $M_Z^2$. The
one--loop and two--loop corrections to the lightest Higgs boson
mass were calculated and analysed in \cite{9} and \cite{10},
respectively. The upper bound on the lightest Higgs boson mass
grows with increasing of $\tan\beta$ and $\ln(M_S^2/m_t^2)$, and
for large $\tan\beta$ ($\tan\beta\gg 1$) reaches
$125-128\text{~GeV}$. In \cite{6} the bounds on the mass of the
Higgs boson in the SM and the MSSM were compared.

However, the large values of $\tan\beta$ are unpreferable for two
reasons. First of them is the proton decay. If one assumes that
electroweak and strong interactions are embedded in the SUSY Grand
Unified Theory (GUT) at high energies then too fast proton decay
is induced due to $d=5$ operators. The major decay mode is $p\to
\bar{\nu}K^{+}$. The proton life--time $\tau_p$ in this case is
inversely proportional to $\tan^2\beta$. When $\tan\beta$ is large
enough, calculated in the framework of SUSY GUT models proton
life--time
 contradicts an experimental restriction on it. Another problem of large $\tan\beta$
scenario concerns flavour--changing neutral currents. The
branching ratio $b\to s\gamma$ rises with $\tan\beta$ as $Br(b\to
s\gamma)\sim \tan^2\beta$. Thus the values of $\tan\beta\gg 1$
($\tan\beta\gtrsim 40$) lead to the unacceptable large
flavour--changing transitions.

In the case of moderate values of $\tan\beta$ ($\tan\beta\lesssim
5$) the $b$--quark and $\tau$--lepton Yukawa couplings are small
and one can get an analytical solution of one--loop
renormalization group equations \cite{11}. For the top quark
Yukawa coupling constant $h_t(t)$ and gauge couplings
 $g_i(t)$ the analytical solution has the following form:
\begin{equation}
\begin{gathered}
Y_t(t)=\frac{\dfrac{E(t)}{6F(t)}}{\left(1+ \dfrac{1}{6Y_t(0)F(t)}
\right)},\qquad
\tilde{\alpha}_i(t)=\frac{\tilde{\alpha}_i(0)}{1+b_i\tilde{\alpha}_i(0)t},\\
E(t)=\left[\frac{\tilde{\alpha}_3(t)}{\tilde{\alpha}_3(0)}\right]^{16/9}
\left[\frac{\tilde{\alpha}_2(t)}{\tilde{\alpha}_2(0)}\right]^{-3}
\left[\frac{\tilde{\alpha}_1(t)}{\tilde{\alpha}_1(0)}\right]^{-13/99},\qquad
F(t)=\int\limits_0^tE(\tau)d\tau,
\end{gathered} \label{3}
\end{equation} where $Y_t(t)=\left(\dfrac{h_t(t)}{4\pi}\right)^2$
and $\tilde{\alpha}_i(t)=\left(\dfrac{g_i(t)}{4\pi}\right)^2$. The
index $i$ varies from 1 to 3, that corresponds to $U(1)$, $SU(2)$,
and $SU(3)$ gauge couplings. The coefficients
 $b_i$ of one--loop beta functions of $\tilde{\alpha}_i(t)$ are $b_1=33/5$, $b_2=1$,
$b_3=-3$. The initial conditions $Y_t(0)$ and
$\tilde{\alpha}_i(0)$ for the MSSM renormalization group equations
are usually set at the Grand Unification scale $M_X\approx 3\times
10^{16}\text{~GeV}$
 where all gauge coupling constants coincide. The variable $t$ is defined by a common way $t=\ln(M_X^2/q^2)$.

Substituting the numerical values of the gauge couplings one finds
that at the electroweak scale the second term in the denominator
of the expression describing the evolution of $Y_t(t)$ is
approximately equal to $\dfrac{1}{10h_t^2(0)}$. When $h_t^2(0)\ge
1$ the dependence of the top quark Yukawa coupling on its initial
value $Y_t(0)$ disappears and all solutions are concentrated near
the quasi--fixed point \cite{12}:
\begin{equation}
\label{4} Y_{\text{QFP}}(t_0)=\frac{E(t_0)}{6F(t_0)},
\end{equation}
where $t_0=2\ln(M_X/M_t^{\text{pole}})\approx 65$. Together with
$Y_t(t)$ the trilinear scalar coupling of the Higgs boson doublet
$H_2$ with stops $A_t(t)$ and some combination of their masses
$\mathfrak{M}_t^2(t)=m_Q^2+m_U^2+m_2^2$ are also driven to the
infrared quasi--fixed points. In the vicinity of these points
$A_t(t)$ is proportional to the universal gaugino mass $M_{1/2}$
at the scale $M_X$ and $\mathfrak{M}_t^2(t)\sim M_{1/2}^2$.
Although the solutions of the MSSM renormalization group equations
achieve quasi--fixed points only for infinite values of $Y_t(0)$,
the deviations from them at the electroweak scale are determined
by the ratio $\dfrac{1}{6F(t_0)Y_t(0)}$ which is quite small if
$h_t^2(0)\ge 1$.

The behaviour of solutions of the MSSM renormalization group
equations near the quasi--fixed point at $\tan\beta\sim 1$ and
particle spectrum have been studied by many authors
[\citen{13}--\citen{15}]. It has been shown that in the vicinity
of this point the $b-\tau$ Yukawa coupling unification is realised
\cite{13}. In the recent publications (see
[\citen{15}--\citen{17}]) the value of $\tan\beta$ that
corresponds to quasi--fixed point regime has been calculated. It
is restricted between 1.3 and 1.8. Such comparatively low values
of $\tan\beta$ yield more stringent bound on the lightest Higgs
boson mass in the MSSM. So, it does not exceed $94\pm
5\text{~GeV}$ [\citen{15},\citen{16}]. The obtained theoretical
bound on $m_h$ has to be compared with the lower experimental one
in the SM since it was computed for the SUSY breaking scale $M_S$
of order of $1\text{~TeV}$ when all other Higgs bosons and
superparticles are heavy enough. The straightforward comparison
shows that the quasi--fixed point scenario and the considerable
part of the MSSM parameter space are almost excluded by LEP\,II
data that stimulates the theoretical analysis of the Higgs sector
in nonminimal supersymmetric models.

In this article the spectrum of the Higgs bosons in the
Next--to--Minimal Supersymmetric Model is reviewed. The lightest
Higgs boson mass in the NMSSM attains its maximum value in the
strong Yukawa coupling limit, when the Yukawa couplings are much
larger then the gauge ones. All supersymmetric models contain a
large number of free parameters what is the main obstacle in the
way of their investigations. For example, each SUSY model includes
three or four independent SUSY breaking constants which determine
the SUSY particles spectrum. Nevertheless, in the strong Yukawa
coupling limit the solutions of the renormalization group
equations are focused near the quasi--fixed point which simplifies
the analysis. We propose a modification of the NMSSM which allows
one to get $m_h\sim 125\text{~GeV}$ for moderate values of
$\tan\beta$ and study Higgs boson spectrum of the model. In the
last part the lightest Higgs boson mass in more complicated SUSY
models is considered.

\section{Higgs sector of the NMSSM}

\subsection{The $\mu$--problem and parameters of the NMSSM}

The simplest extension of the MSSM is the Next--to--Minimal
Supersymmetric Standard Model (NMSSM). Historically the NMSSM was
suggested as a solution of the $\mu$--problem in the supergravity
(SUGRA) models \cite{18}. In addition to observable superfields
these models contain a "hidden" sector where local supersymmetry
is broken. In the superstring inspired SUGRA models the "hidden"
sector always includes the singlet dilaton $S$ and moduli $T_m$
superfields. They appear in the four--dimensional theory as a
result of compactification of extra dimensions. The full
superpotential of SUGRA models can be presented as an expansion in
powers of observable superfields
\begin{equation}
W=\hat{W}_0(S,T_m)+\mu(S,T_m)(\hat{H}_1\hat{H}_2)+h_t(S,T_i)(\hat{Q}\hat{H}_2)\hat{U}_R^c+
\dots, \label{5}
\end{equation}
where $\hat{W}_0(S,T_m)$ is the superpotential of the "hidden"
sector. From the expansion (\ref{5}) it is obvious that parameter
$\mu$ should be of order of the Planck scale because that is the
only scale characterising the "hidden" (gravity) sector of the
theory. On the other hand, if $\mu\sim M_{\text{Pl}}$ then the
Higgs doublets get huge positive masses $m_{H_1,H_2}^2\simeq
\mu^2\simeq M_{\text{Pl}}^2$ and electroweak symmetry breaking
does not occur at all.

In the NMSSM a new singlet superfield $Y$ is introduced. By
definition the superpotential of this model is invariant with
respect to the $Z_3$ discrete transformations \cite{19}. The $Z_3$
symmetry usually arises in the superstring inspired models in
which all observable superfields are massless in the exact
supersymmetry limit. The term $\mu(\hat{H}_1\hat{H}_2)$ does not
satisfy the last requirement. Therefore it must be eliminated from
the NMSSM superpotential. Instead of it the sum of two terms
\begin{equation}
W_h=\lambda\hat{Y}(\hat{H}_1\hat{H}_2)+\frac{\varkappa}{3}\hat{Y}^3
\label{6}
\end{equation}
arises [\citen{18}--\citen{20}]. After electroweak symmetry
breaking the singlet field $Y$ acquires a nonzero vacuum
expectation value ($\langle Y\rangle=y/\sqrt{2}$) and the
effective $\mu$--term ($\mu=\lambda y/\sqrt{2}$) is generated.

The NMSSM superpotential contains a lot of Yukawa couplings. But
at the moderate values of $\tan\beta$ all of them are small and
can be neglected except for $h_t$, $\lambda$, and $\varkappa$. In
addition to the Yukawa couplings the lagrangian of the NMSSM
contains a large number of soft supersymmetry breaking parameters.
Each of the scalar and gaugino fields has a soft mass ($m_i$ and
$M_i$, respectively). Each of the Yukawa couplings corresponds to
the trilinear scalar coupling $A_i$ in the full lagrangian. The
number of these unknown parameters can be considerably reduced if
one assumes the universality of the soft SUSY breaking terms at
the scale $M_X$. Thus, only three independent dimensional
parameters are left: the universal gaugino mass $M_{1/2}$, the
universal scalar mass $m_0$, and the universal trilinear coupling
of scalar fields $A$. Naturally universal soft SUSY breaking terms
appear in the minimal supergravity model \cite{21} and in the
simplest models deduced from the superstring theories \cite{22}.
The universal parameters of soft supersymmetry breaking determined
at the Grand Unification scale have to be considered as boundary
conditions for the renormalization group equations that describe
the evolution of all fundamental couplings up to electroweak scale
or SUSY breaking scale. The complete system of the NMSSM
renormalization group equations can be found in
[\citen{23},\citen{24}].

\subsection{The CP--even Higgs boson spectrum}

The Higgs sector of the Next-to-Minimal Supersymmetric Standard
Model includes six massive states. Three of them are CP--even
fields, two are CP--odd fields, and one is a charged field. The
determinants of the mass matrices of the CP--odd and charged Higgs
bosons equal zero. It corresponds to the appearance of two
Goldstone bosons:
\begin{equation}
\begin{aligned}
\eta^0&=\sqrt{2}\sin\beta\,\iim H_2^0+\sqrt{2}\cos\beta\,\iim
H_1^0
\\ \eta^+&=\sin\beta\,H_2^++\cos\beta\,(H_1^-)^*
\end{aligned}\label{7}
\end{equation}
which are swallowed up by the massive vector $W^{\pm}$ and
$Z$--bosons during the spontaneous breaking of $SU(2)\otimes U(1)$
symmetry. For this reason the masses of neutral CP--odd bosons and
charged boson are easily calculated.

In the CP--even Higgs sector the situation is more complex. The
CP--even states arise as a result of mixing of the real parts of
the neutral components of two Higgs doublets with the real part of
the field $Y$. The determinant of their mass matrix does not
vanish and in order to calculate its eigenvalues one has to
diagonalize the ($3\times 3$) mass matrix. Instead of $\rre
H_1^0$, $\rre H_2^0$, and $\rre Y$ it is much more convenient to
consider their linear combinations:
\begin{equation}
\begin{aligned}
\chi_1&=\sqrt{2}\cos\beta\rre H_1^0+\sqrt{2}\sin\beta\rre H_2^0,
\\ \chi_2&=-\sqrt{2}\sin\beta\rre H_1^0+\sqrt{2}\cos\beta\rre
H_2^0\, ,\\ \chi_3&=\sqrt{2}\rre Y.
\end{aligned}
\label{8}
\end{equation}
In the basis (\ref{8}) the mass matrix of CP--even Higgs fields
can be simply written in the following symmetrical form (see
\cite{25}):

\begin{equation}
M^2_{ij}=\begin{pmatrix} \dfrac{\partial^2 V}{\partial v^2} &
\dfrac{1}{v}\dfrac{\partial^2 V}{\partial v\partial\beta} &
\dfrac{\partial^2 V}{\partial v\partial y} \\
\dfrac{1}{v}\dfrac{\partial^2 V}{\partial v\partial\beta} &
\dfrac{1}{v^2}\dfrac{\partial^2 V}{\partial\beta^2} &
\dfrac{1}{v}\dfrac{\partial^2 V}{\partial y\partial\beta} \\
\dfrac{\partial^2 V}{\partial v\partial y} &
\dfrac{1}{v}\dfrac{\partial^2 V}{\partial y\partial\beta} &
\dfrac{\partial^2 V}{\partial y^2}
\end{pmatrix},
\label{9}
\end{equation}
where $V(v_1, v_2, y)$ is the effective potential of the NMSSM
Higgs sector:
\begin{equation}
\begin{gathered}
V(H_1,H_2,Y)=m_1^2|H_1|^2+m_2^2|H_2|^2+m_y^2|Y|^2+{}\\
{}+\left(\lambda A_\lambda(H_1H_2)Y+\frac{\varkappa}{3}A_\varkappa
Y^3+\lambda\varkappa(H_1H_2)(Y^*)^2+\mathrm{h.c.}\right)+{}\\
{}+\lambda^2|h_1H_2|^2+\lambda^2|H_2|^2|Y|^2+\lambda^2|H_1|^2|Y|^2+\varkappa^2|Y|^4+{}\\
{}+\frac{{g'}^2}{8}(|H_1|^2-|H_2|^2)^2+\frac{g^2}{8}(H_1^+\boldsymbol{\sigma}H_1+H_2^+\boldsymbol{\sigma}H_2)^2+\Delta
V,
\end{gathered}
\label{Higgs-NMSSM}
\end{equation}
where $\Delta V(H_1, H_2, Y)$ is the sum of loop corrections to
the effective potential, $g$ and $g'$ are the gauge constants of
the $SU(2)$ and $U(1)$ interactions, respectively
($g_1=\sqrt{5/3}\,g'$).

It is well known that {\itshape the minimum eigenvalue of a matrix
does not exceed its minimum diagonal element.} Thus the lightest
CP--even Higgs boson mass is always smaller than
\begin{equation}
m_h^2\le M^2_{11}=\frac{\partial^2 V}{\partial v^2}=
\frac{\lambda^2}{2}v^2\sin^2 2\beta+M_Z^2\cos^2 2\beta+\Delta\, .
\label{10}
\end{equation}
In the right--hand side of inequality (\ref{10}) $\Delta$ is the
contribution of loop corrections to the Higgs boson potential. The
expression (\ref{10}) was obtained in the tree level approximation
($\Delta=0$) in \cite{20}. The contribution of loop corrections to
the upper bound on the lightest Higgs boson mass in the NMSSM is
almost the same as in the minimal SUSY model. In particular, in
order to calculate the corrections from the $t$--quark and its
superpartners one has to replace the parameter $\mu$ in the
corresponding formulas of the MSSM by $\lambda y/\sqrt{2}$. The
Higgs boson sector of the NMSSM and loop corrections to it were
studied in [\citen{24}--\citen{26}]. Let us also remark that the
upper bound on $m_h$ in the NMSSM was compared (see \cite{7}) with
theoretical bounds in the SM and in its minimal supersymmetric
extension.

The calculation of CP--even Higgs spectrum is simplified in the
most interesting realistic case when all superparticles are heavy
($M_S\gg M_Z$). In this case the contributions of new particles to
the electroweak observables are suppressed as
$\left(\dfrac{M_Z}{M_S}\right)^2$ (see, for example, \cite{27}).
On the other hand, the prediction for the values of strong
coupling constant at the electroweak scale $\alpha_3(M_Z)$ that
can be obtained from the gauge coupling unification \cite{28} is
improved with increasing of the supersymmetry breaking scale
$M_S$. For $M_S\simeq 1\text{~TeV}$ it becomes close to the
$\alpha_3(M_Z)=0.118(3)$ which has been found independently from
the analysis of the experimental data \cite{29}. Also it should be
noted that the lightest Higgs boson mass reaches its maximal value
in the SUSY models for $M_S\sim 1-3\text{~TeV}$.

In the considered limit the mass matrix (\ref{9}) has a
hierarchical structure and can be represented as a sum of two
matrices \cite{25}:
\begin{equation}
M^2_{ij}=\begin{pmatrix} E^2_1 & 0 & 0\\ 0 & E^2_2 & 0\\ 0 & 0 &
E^2_3 \end{pmatrix}+\begin{pmatrix} V_{11} & V_{12} & V_{13}\\
V_{21} & V_{22} & V_{23}\\ V_{31} & V_{32} & V_{33} \end{pmatrix}.
\label{11}
\end{equation}
The first matrix is diagonal with $E_1^2=0$ and $E_{2,3}^2\sim
M_S^2$. The matrix elements $V_{11}$, $V_{22}$, $V_{33}$,
$V_{12}=V_{21}$ are of order of $M_Z^2$. The other matrix elements
that correspond to mixing of $\chi_1$ and $\chi_2$ with $\chi_3$
are equal to
\begin{equation}
V_{13}=V_{31}=\lambda v X_1,\qquad V_{23}=V_{32}=\lambda v X_2,
\label{12} \end{equation} where $X_1\sim X_2\sim M_S$.

Considering the ratio $\dfrac{M_Z^2}{E^2_{2,3}}$ as a small
parameter it is easy to diagonalize the mass matrix (\ref{11}) by
means of usual quantum mechanical perturbation theory. For the
Higgs boson masses it yields:
\begin{equation}
\begin{aligned}
m_S^2&\approx E_3^2+V_{33}+\lambda^2
v^2\frac{X_1^2}{E_3^2}+\lambda^2 v^2 \frac{X_2^2}{E_3^2-E_2^2},\\
m_H^2&\approx E_2^2+V_{22}+\lambda^2
v^2\frac{X_2^2}{E_2^2-E_3^2},\\ m_h^2&\approx
\frac{\lambda^2}{2}v^2\sin^2 2\beta+ M_Z^2\cos^2
2\beta+\Delta-\lambda^2 v^2\frac{X_1^2}{E_3^2}.
\end{aligned}
\label{13}
\end{equation}
The explicit expressions for the $E_i^2$ and $V_{ij}$ can be found
in \cite{25}. For simplicity we restrict our consideration to the
first order of perturbation theory and neglect matrix element
$V_{12}$ because its contribution to $m_i^2$ is of order of
$\dfrac{M_Z^4}{M_S^2}$.

The perturbation theory becomes inapplicable when
$|E_2^2-E_3^2|\sim \lambda v X_2$. However, the mass matrix
(\ref{11}) can be easily diagonalised even in this case. In order
to do this one should choose the basis where the matrix element
$M_{23}$ is zero. After that in the new basis the Higgs boson
masses can be computed using ordinary perturbation theory.

The first three terms in the last relation in (\ref{13}) reproduce
the upper bound on the lightest Higgs boson mass in the NMSSM.
Their sum is equal to $V_{11}$ in our notations. The last term in
this expression gives a negative contribution to $m_h$. Even when
the ratio $\dfrac{M^2_Z}{M_S^2}$ goes to zero it does not vanish.
Thus in the NMSSM the mass of the lightest CP--even Higgs boson
can be considerably less than its upper bound \cite{25}.

\subsection{Renormalization of the Yukawa couplings and soft SUSY breaking terms}

According to inequality (\ref{10}) the upper bound on $m_h$ rises
when $\lambda$ increases and the value of $\tan\beta$ diminishes.
For $\tan\beta\gg 1$ the value of $\sin2\beta$ goes to zero and
the upper bound on the lightest Higgs boson mass in the NMSSM
coincides with that one in the minimal SUSY model. With decreasing
of $\tan\beta$ the top quark Yukawa coupling at the electroweak
scale $h_t(t_0)$ grows. The analysis of the solutions of the NMSSM
renormalization group equations reveals that a rise of the values
of the Yukawa couplings at the electroweak scale entails an
increase of them at the Grand Unification scale. As a result the
upper bound on the lightest Higgs boson mass in the NMSSM reaches
its maximum value in the strong Yukawa coupling limit when the
Yukawa couplings are much larger than the gauge ones at the scale
$M_X$.

The renormalization of the NMSSM coupling constants in the strong
Yukawa coupling limit has been studied in [\citen{30},\citen{31}].
In the considered case the Yukawa couplings are attracted towards
a Hill type effective (quasi--fixed) line ($\varkappa=0$) or
surface ($\varkappa\ne 0$) which restrict the allowed regions of
$h_t$, $\lambda$, and $\varkappa$. Outside this range the
solutions of renormalization group equations blow up before the
Grand Unification scale $M_X$ and perturbation theory becomes
invalid at $q^2\sim M_X^2$. While the values of the Yukawa
couplings at the scale $M_X$ grow, the region, where all solutions
are concentrated, shrinks and $h^2_t(0)$, $\lambda^2(0)$,
$\varkappa^2(0)$ are focused near the quasi--fixed points
\cite{30}. These points appear as a result of intersection of the
quasi--fixed line or surface with the invariant (fixed) line. The
latter connects the stable fixed point in the strong Yukawa
coupling regime \cite{32} with infrared fixed point of the NMSSM
renormalization group equations \cite{33}. The properties of
invariant lines and surfaces were reviewed in detail in
[\citen{5},\citen{34}].

When the values of Yukawa couplings tend to quasi--fixed points
the trilinear scalar couplings $A_i(t)$ and some combinations of
scalar particle masses $\mathfrak{M}_i^2(t)$, where
\begin{equation}
\begin{aligned}
\mathfrak{M}_t^2(t)&=m_2^2(t)+m_Q^2(t)+m_U^2(t), \\
\mathfrak{M}_{\lambda}^2(t)&=m_1^2(t)+m_2^2(t)+m_y^2(t), \\
\mathfrak{M}_{\varkappa}^2(t)&=3m_y^2(t),
\end{aligned}
\label{14}
\end{equation}
become insensitive to their initial values $A$ and $3m_0^2$ at the
scale $M_X$ \cite{31}. For the universal boundary conditions one
has
\begin{equation}
\begin{gathered}
A_i(t)=e_i(t)A+f_i(t)M_{1/2}, \\
m_i^2(t)=a_i(t)m_0^2+b_i(t)M_{1/2}^2+c_i(t)A M_{1/2}+d_i(t)A^2.
\end{gathered}
\label{15}
\end{equation}
The functions $e_i(t)$, $f_i(t)$, $a_i(t)$, $b_i(t)$, $c_i(t)$,
and $d_i(t)$ remain unknown since an analytical solution of the
NMSSM renormalization group equations has not been found yet.
While the Yukawa couplings tend to infinity the values of
functions $e_i(t_0)$, $c_i(t_0)$, and $d_i(t_0)$ vanish. It means
the solutions of renormalization group equations go to the
quasi--fixed points too. In the vicinity of the quasi--fixed
points $A_i(t)$ are proportional to $M_{1/2}$ and
$\mathfrak{M}_i^2(t)\sim M_{1/2}^2$.

\section{Particle spectrum in the modified NMSSM}

\subsection{The modified NMSSM}

The fundamental parameters of the NMSSM at the Grand Unification
scale have to be adjusted so that the minimisation conditions of
the effective Higgs boson potential are satisfied:
\begin{equation}
\frac{\partial V(v_1,v_2,y)}{\partial v_1}=0,\quad \frac{\partial
V(v_1,v_2,y)}{\partial v_2}=0,\quad \frac{\partial
V(v_1,v_2,y)}{\partial y}=0. \label{16}
\end{equation}
Since the vacuum expectation value $v$ is known they can be used
for the calculation of $A$, $m_0$, and $M_{1/2}$. But in the
strong Yukawa coupling limit it is impossible to get the real
solution of nonlinear algebraic equations (\ref{16}). Thus
although the recent investigations [\citen{35},\citen{36}] reveal
that the upper bound on the lightest Higgs boson mass in the NMSSM
is larger than the one in the MSSM by $7-10\text{~GeV}$, in the
considered region of the NMSSM parameter space the
self--consistent solution cannot be obtained. Such solution of
equations (\ref{16}) appears for $\lambda^2(0), \varkappa^2(0)
\leq 0.1$ when the upper bound on $m_h$ is the same as in the
minimal supersymmetric model.

Moreover due to the $Z_3$ symmetry three degenerate vacuum
configurations arise. After a phase transition at the electroweak
scale the Universe is filled by three degenerate phases. The
regions with different phases are separated from each other by
domain walls. The cosmological observations do not confirm the
existence of such domain walls. The domain structure of the vacuum
is destroyed if the discrete $Z_3$ symmetry of the NMSSM
lagrangian disappears. An attempt of $Z_3$ symmetry breaking by
means of operators of dimension $d=5$ was made in \cite{37}. It
was shown that their introduction leads to quadratic divergences
in the two--loop approximation, i.e., to the appearing of
hierarchy problem. As a result the vacuum expectation value of $Y$
turns out to be of the order of $10^{11}\text{~GeV}$.

Thus, in order to avoid the domain wall problem and get the
self--consistent solution in the strong Yukawa coupling limit, one
has to modify the NMSSM. The simplest way is to introduce the
bilinear terms $\mu(\hat{H}_1\hat{H}_2)$ and $\mu'\hat{Y}^2$ in
the superpotential which are not forbidden by electroweak
symmetry. At the same time one can omit the coupling $\varkappa$,
that allows to simplify the analysis of the modified NMSSM. In
this case ($\varkappa=0$) the upper bound on the lightest Higgs
boson mass reaches its maximum value. Neglecting all the Yukawa
constants except for $h_t$ and $\lambda$ one gets the following
expression for the modified NMSSM (MNSSM) superpotential
\cite{38}:
\begin{equation}
W_{\text{MNSSM}}=\mu(\hat{H}_1\hat{H}_2)+\mu'\hat{Y}^2+\lambda
\hat{Y}(\hat{H}_1\hat{H}_2)+ h_t(\hat{H}_2\hat{Q})\hat{U}^C_R.
\label{17}
\end{equation}
The bilinear terms in the superpotential (\ref{17}) break the
$Z_3$ symmetry and the domain walls do not arise, because the
degenerated vacua do not exist. The introduction of the parameter
$\mu$ permits to obtain the self--consistent solution of algebraic
equations (\ref{16}) when $h_t^2(0), \lambda^2(0)\gg g_i^2(0)$. In
the supergravity models the bilinear terms may be generated due to
the additional terms $[Z(H_1 H_2)+Z'Y^2+\mathrm{h.c.}]$ in the
K\"ahller potential [\citen{39},\citen{40}] or due to
nonrenormalizable interaction of the Higgs doublet superfields
with the "hidden" sector ones [\citen{40},\citen{41}].

The effective Higgs boson potential of MNSSM can be written as the
sum
\begin{equation}
\begin{gathered}
V(H_1,H_2,Y)=\mu_1^2|H_1|^2+\mu_2^2|H_2|^2+\mu_y^2|Y|^2+\Biggl[\mu_3^2(H_1H_2)+
\mu_4^2Y^2+{}\\ {}+\lambda A_\lambda
Y(H_1H_2)+\lambda\mu'Y^{*}(H_1H_2)+\lambda\mu
Y\Biggl(|H_1|^2+|H_2|^2\Biggr)+\mathrm{h.c.}\Biggr]+{}\\
{}+\lambda^2|(H_1H_2)|^2+\lambda^2Y^2\Biggl(|H_1|^2+|H_2|^2\Biggr)+
\frac{{g'}^2}{8}\Biggl(|H_2|^2-|H_1|^2\Biggr)^2+{}\\
{}+\frac{g^2}{8}\Biggl(H_1^+\boldsymbol{\sigma}
H_1+H_2^+\boldsymbol{\sigma} H_2\Biggr)^2+ \Delta V(H_1,H_2,Y),
\end{gathered}
\label{18}
\end{equation}
where the parameters $\mu_i^2$ are expressed via the soft SUSY
breaking terms as follows:
\begin{gather*}
 \mu_1^2=m_1^2+\mu^2, \qquad
\mu_2^2=m_2^2+\mu^2, \qquad \mu_y^2=m_y^2+{\mu'}^2, \\
\mu_3^2=B\mu, \qquad \mu_4^2=\frac{1}{2}B'\mu'.
\end{gather*}
The $\mu$--terms in the superpotential (\ref{17}) lead to the
appearance of bilinear scalar couplings $B$ and $B'$ in the
effective potential of the Higgs fields. They arise as a result of
the soft supersymmetry breaking. The values of $B$ and $B'$ depend
on the mechanism of the $\mu$ and $\mu'$ generation. For a minimal
choice of the fundamental parameters all soft scalar masses and
all bilinear scalar couplings should be put equal at the scale
$M_X$:
\begin{gather*} m_1^2(M_X)=m_2^2(M_X)=m_y^2(M_X)=m_0^2, \\
B(M_X)=B'(M_X)=B_0. \end{gather*}
Thus in addition to the
parameters of the SM the modified NMSSM contains seven independent
ones:
\[ \lambda,\mu,\mu',A,B_0,m_0^2,M_{1/2}. \]

\subsection{The procedure of the analysis}

Although the parameter space of the considered model is enlarged
it is possible to get some predictions for the Higgs boson
spectrum in the strong Yukawa coupling limit. It is reasonable to
start our analysis from the quasi--fixed point of MNSSM \cite{30}:
\begin{equation}
\begin{aligned}
\rho^{\text{QFP}}_t(t_0)&=0.803,\quad
&\rho^{\text{QFP}}_{A_t}(t_0)&=1.77,\quad
&\rho^{\text{QFP}}_{\mathfrak{M}_t^2}(t_0)&=6.09,\\
\rho^{\text{QFP}}_{\lambda}(t_0)&=0.224,\quad
&\rho^{\text{QFP}}_{A_{\lambda}}(t_0)&=-0.42,\quad
&\rho^{\text{QFP}}_{\mathfrak{M}_{\lambda}^2}(t_0)&=-2.28,
\end{aligned}
\label{19}
\end{equation}
because all solutions of renormalization group equations are
concentrated in the vicinity of it at $h_t^2(0)$, $\lambda^2(0)\gg
g^2_i(0)$. In the relations (\ref{19}) the following notations are
used:
$\rho_{t,\lambda}(t)=\dfrac{Y_{t,\lambda}(t)}{\tilde{\alpha}_3(t)}$,
$Y_{t}=\dfrac{h_t^2(t)}{(4\pi)^2}$,
$Y_{\lambda}(t)=\dfrac{\lambda^2(t)}{(4\pi)^2}$,
$\rho_{A_{t,\lambda}}=\dfrac{A_{t,\lambda}}{M_{1/2}}$, and
$\rho_{\mathfrak{M}^2_{t,\lambda}}=\dfrac{\mathfrak{M}^2_{t,\lambda}}{M^2_{1/2}}$.

For given $\rho_t(t_0)$ (\ref{19}) the value of $\tan\beta$ can be
extracted from the expression which relates the running $t$--quark
mass $m_t(M_t^{\text{pole}})$ with $h_t(t_0)$:
\begin{equation}
m_t(M_t^{\text{pole}})=\frac{1}{\sqrt{2}}h_t(M_t^{\text{pole}})v\sin\beta.
\label{20}
\end{equation}
We substitute in the left side of equation (\ref{20}) the value of
the running top quark mass $m_t(M_t^{\text{pole}})=165\pm
5\text{~GeV}$ calculating in the $\overline{MS}$ scheme \cite{42}.
The uncertainty in the $m_t(M_t^{\text{pole}})$ is determined by
the experimental error with which the
 top quark pole mass is measured: $M_t^{\text{pole}}=174.3\pm 5.1\text{~GeV}$ \cite{43}.
In the infrared quasi--fixed point regime that corresponds to
maximal allowed values of $h_t^2(0)$ and $\lambda^2(0)$ (we take
$h_t^2(0)=\lambda^2(0)=10$) we obtain $\tan\beta\approx 1.88$ for
$m_t(M_t^{\text{pole}})=165\text{~GeV}$ \cite{38}.

At the first step of our analysis the supersymmetry breaking scale
is also fixed by means of the condition
$M_3(1000\text{~GeV})=1000\text{~GeV}$, where $M_3$ is the gluino
mass. This condition permits to calculate immediately the
universal gaugino mass at the Grand Unification scale. It ensures
that the superparticles are much heavier than the observable ones.

Next we use the equations (\ref{16}) that define the minimum of
the Higgs boson potential of the modified NMSSM to restrict the
allowed region of the parameter space. Instead of $\mu$ it is more
convenient to introduce $\mu_{\text{eff}}=\mu+\lambda y/\sqrt{2}$.
Then after some transformations we obtain:
\begin{equation}
\label{21}
\begin{gathered}
\mu_{\text{eff}}^2=\frac{m_1^2-m_2^2\tan^2\beta+\Delta_Z}{\tan^2\beta-1}
 -\frac{1}{2}M_Z^2\\
\sin 2\beta=\frac{-2\left(B\mu
 +\dfrac{\lambda y X_2}{\cos 2\beta}\right)}{m_1^2+m_2^2+2\mu_{\text{eff}}^2+\frac{\lambda^2}{2}v^2
 +\Delta_\beta}\\
y\left(m_y^2+{\mu'}^2+B'\mu'\right)={\frac{\lambda}{2}v^2 X_1-
\Delta_y},
\end{gathered}
\end{equation}
where $\Delta_i$ are the contributions of loop corrections and \[
X_1=\frac{1}{\sqrt{2}}\left(2\mu_{\text{eff}}+(\mu'+A_{\lambda})
\sin 2\beta\right),\qquad
X_2=\frac{1}{\sqrt{2}}\left(\mu'+A_{\lambda}\right)\cos 2\beta.
\]
As the values of $v$ and $\tan\beta$ are known one can find from
the equations (\ref{21}) the vacuum expectation value $y$ and
parameters $B_0$ and $\mu_{\text{eff}}$. In the numerical analysis
we take into account only the loop corrections from $t$--quark and
its superpartners, because they give a leading contribution.
Therefore $\Delta_i$ are the functions of $\mu_{\text{eff}}$ and
do not depend on $B_0$ and $y$. From the first equation of
(\ref{21}) the absolute value of $\mu_{\text{eff}}$ is calculated.
The sign of $\mu_{\text{eff}}$ is not determined and should be
considered as a free parameter. The bilinear scalar coupling $B_0$
and vacuum expectation value $y$ are computed from the two other
equations of (\ref{21}). The last of them points out that the
value of $y$ is of the order of $\lambda v^2/M_S$ and much smaller
than $v$ if the superparticles are heavy enough.

Since $\mu_{\text{eff}}$, $B_0$, and $y$ have been found we
investigate the dependence of the Higgs boson spectrum on $A$,
$m_0$, and $\mu'$ using the relations (\ref{15}). For the masses
of CP--odd states one can get an exact analytical result:
\[
m_{A_1,A_2}^2=\frac{1}{2}\left(m_A^2+m_B^2\pm\sqrt{
\left(m_A^2-m_B^2\right)^2+ 4\left(\frac{\lambda v}{\sqrt{2}}
(\mu'+A_\lambda)+\Delta_0 \right)^2}\right),
\]
\begin{gather}
m_A^2=m_1^2+m_2^2+2\mu_{\text{eff}}^2+\frac{\lambda^2}{2}v^2+
\Delta_A, \label{22} \\
m_B^2=m_y^2+{\mu'}^2-B'\mu'+\frac{\lambda^2}{2}v^2+ \Delta_3.
\nonumber
\end{gather}

The mass matrix of CP--even Higgs sector has a hierarchical
structure and can be written in the form (\ref{11}) with
\begin{equation}
\begin{aligned}
E_2^2&=m_1^2+m_2^2+2\mu_{\text{eff}}^2, \qquad
E_3^2=m_y^2+{\mu'}^2+B'\mu',
\\
V_{11}&=M_Z^2\cos^2 2\beta+\frac{1}{2}\lambda^2 v^2\sin^2 2\beta+
\Delta_{11},
\\
V_{12}&=V_{21}=\left(\frac{1}{4}\lambda^2
v^2-\frac{1}{2}M_Z^2\right)\sin 4\beta+\Delta_{12}, \\
V_{22}&=M_Z^2\sin^2 2\beta+\frac{1}{2}\lambda^2 v^2 \cos^2
2\beta+\Delta_A+\Delta_{22}, \\ V_{13}&=V_{31}=\lambda v
X_1+\Delta_{13}, \qquad V_{23}=V_{32}=\lambda v X_2+\Delta_{23},
\\ V_{33}&=\frac{1}{2}\lambda^2 v^2+\Delta_{33}.
\end{aligned}
\label{23}
\end{equation}
In the formulas (\ref{22}) and (\ref{23}) $\Delta_0, \Delta_3,
\Delta_A$ and $\Delta_{ij}$ ($\Delta_{11}=\Delta$) are the loop
corrections to the CP--odd and CP--even mass matrices. The mass
matrix of  CP--even Higgs sector can be diagonalised and the
expressions for the
 masses of CP--odd states (\ref{22}) can be simplified using the perturbation theory of
 quantum mechanics. In the main order of perturbation theory the masses of the heavy Higgs bosons are:
$m_H^2\approx E_2^2$, $m_S^2\approx E_3^2$, $m_{A_1}^2\approx
m_B^2$, and $m_{A_2}^2\approx m_A^2$, while the first order
perturbation corrections have the form similar to given in
Eq.(\ref{13}).

\subsection{Numerical results}

The results of the numerical analysis of the particle spectrum near the MNSSM quasi--fixed
point are presented in Figs.1--3. There are two regions of the MNSSM parameter space.
In one of them the mass of the lightest CP--even Higgs boson is larger than the upper
bound on $m_h$ in the MSSM (see Figs. 1a and 2a) whereas in the other region it is smaller
(see Figs. 1b and 2b). As follows from the relations (\ref{13}) the lightest Higgs boson mass
in the NMSSM and in its modification attains its upper bound when $V_{13}$ (or $X_1$) goes to zero.
In the MNSSM it happens if
\begin{equation}
\mu'=-\frac{2\mu_{\text{eff}}}{\sin 2\beta}-A_{\lambda}-
\frac{\sqrt{2}\Delta_{13}}{\lambda v\sin2\beta}. \label{24}
\end{equation}
Thus $m_h$ is larger in that part of the parameter
space where the signs of $\mu$ and $\mu'$ are opposite. If $\mu'$
tends to infinity then the singlet CP--odd and CP--even fields get
huge masses and their contribution to the effective potential of
the Higgs bosons vanishes due to the decoupling property. In the
considered limit the lightest Higgs boson mass is the same as in
the minimal SUSY model.

It is necessary to emphasize that the one--loop and even two--loop
corrections give an appreciable contribution to the mass of the
lightest CP--even Higgs boson. So the two--loop corrections
\cite{10} reduce its mass approximately by $10\text{~GeV}$. They
nearly compensate the growth of lightest Higgs boson mass with
increasing of SUSY breaking scale $M_S$ which arises because of
one--loop corrections. Due to loop corrections the values of $m_h$
for $\mu_{\text{eff}}>0$ and $\mu_{\text{eff}}<0$ become
different. The contribution of loop corrections $\Delta$ rises as
the stop mixing parameter $X_t=A_t+\mu_{\text{eff}}/\tan\beta$
increases. Since near the quasi--fixed point $A_t<0$ (see
(\ref{19})), the absolute value of $X_t$ is larger if
$\mu_{\text{eff}}<0$.

As the part of the parameter space where $\mu_{\text{eff}}$ and
$\mu'$ have the same signs is almost excluded by Higgs searches
data from LEP\,II, we investigate the particle spectrum in the
case when the signs of $\mu_{\text{eff}}$ and $\mu'$ are opposite.
In the most interesting region, where the lightest Higgs boson
mass is close to its upper bound, the value of $\mu'$ is
considerably larger than $\mu_{\text{eff}}$ and $M_S$ (see
(\ref{24})). Moreover the product $B'\mu'$ is positive. Indeed
from the second relation of the system (\ref{21}), which defines
the minimum of the MNSSM Higgs boson potential (\ref{18}), it
follows that the bilinear scalar coupling $B$ and $\mu$ have
different signs. As a consequence near the maximum of the curves
in Figs. 1a and 2a the sign of $B'$ coincides with the sign of
$\mu'$.

It means that the heaviest particle in the modified NMSSM is the
CP--even Higgs boson that corresponds to the neutral field $Y$.
Its mass is $m_S^2>{\mu'}^2$ and is substantially larger than the
scale of supersymmetry breaking. As one can see from Figs. 1c and
2c the mass of the other heavy CP--even Higgs boson $m_H$ is
almost insensitive to the value of $\mu'$ since $m_S^2\gg m_H^2$.
The masses of CP--odd states are always smaller than ${\mu'}^2$.
If the value of $\mu'$ grows the mass of the heaviest CP--odd
state $m_{A_1}^2$ increases too as $m_{A_1}^2\sim {\mu'}^2$. When
the value of $\mu'$ diminishes the lightest CP--odd boson mass
$m_{A_2}$ decreases and for $\mu'\sim B'$ becomes of the order of
electroweak scale. At given values of $\mu'$ the low constraint on
$\mu'$ appears which comes from the requirement of $m_{A_2}^2>0$.
However even if the mass of the lightest CP--odd state is of the
order of $M_Z$, it will be quite difficult to observe it in future
experiments because the main contribution to its wave function
gives the CP--odd component of the singlet field $Y$. The heaviest
fermion in the modified NMSSM is the neutralino
($m_{\tilde{\chi}_5}$) which is a superpartner of the scalar field
$Y$. Its mass is proportional to $\mu'$. The remaining masses of
neutralinos ($m_{\tilde{\chi}_i}$), charginos
($m_{\tilde{\chi}^{\pm}_i}$), squarks, and sleptons do not depend
on the $\mu'$.

The spectrum of new fermion states, squarks and sleptons is also
insensitive to the choice of the parameter $A$, since near the
quasi--fixed point (\ref{19}) the dependence of scalar masses
$m_i^2$ and trilinear scalar couplings $A_i$ on it disappears. For
this reason the lightest CP--even Higgs boson mass is almost
independent of $A$. Nevertheless the dependence of the heavy Higgs
boson spectrum on the universal trilinear scalar coupling $A$ is
conserved. It occurs because the bilinear scalar coupling $B'$ is
proportional to $A$. The dependence of the Higgs boson masses on
the parameter $A$ for $m_0=0$ is presented in Figs. 1d and 2d.

Although everywhere in Figs. 1 and 2 we put $m_0=0$, the qualitative pattern of the
 particle spectrum does not change if the universal soft scalar mass varies from zero to $M_{1/2}^2$.
 It should be noted that the masses of squarks, sleptons, heavy Higgs bosons, heavy charginos and
neutralinos rise with increasing of $m_0$ while the spectrum of the lightest particles
remains unchanged.

Up to now the particle spectrum in the quasi--fixed point regime
which corresponds to the initial values of the Yukawa couplings
$h_t^2(0)=\lambda^2(0)=10$ has been studied. In the vicinity of
the quasi--fixed point (\ref{19}) for
$m_t(M_t^{\text{pole}})=165\text{~GeV}$ and $M_3\le 2\text{~TeV}$
the lightest Higgs boson mass does not exceed $127\text{~GeV}$.
The results presented in Table point out that the qualitative
pattern or MNSSM particle spectrum does not change even if
$h_t^2(0)\gg\lambda^2(0)$ or $h_t^2(0)\ll\lambda^2(0)$ as long as
the Yukawa couplings are much larger than the gauge ones at the
scale $M_X$. The masses of the Higgs bosons and the masses of the
superpartners of the observable particles were calculated there
for the values of $\mu'$ computed by the formula (\ref{24}) and
$A=m_0=0$. At the same time from the Table one can see that the
numerical value of the lightest Higgs boson mass is raised from
$105-113\text{~GeV}$ for $\lambda^2(0)=2$ to $118-128\text{~GeV}$
for $\lambda^2(0)=10$.

Therefore at the last stage of our analysis we investigate the
dependence of the upper bound on $m_h$ on the choice of the Yukawa
couplings at the Grand Unification scale. For each $h_t^2(t_0)$ we
find the value of $\tan\beta$ using the relation (\ref{20}) and
choose $\lambda^2(t_0)$ and $\mu'$ so that the lightest Higgs
boson mass attains its upper bound. The obtained curve
$m_h(\tan\beta)$ is plotted in Fig. 3, where the upper bound on
$m_h$ in the MSSM as a function of $\tan\beta$ is also presented.
Two bounds are very close for large $\tan\beta$ ($\tan\beta\gg
1$). The curve $m_h(\tan\beta)$ in the MNSSM reaches its maximum
when $\tan\beta=2.2-2.4$ which corresponds to the strong Yukawa
coupling limit. The numerical analysis reveals that the mass of
the lightest CP--even Higgs boson in the modified NMSSM is always
smaller than $130.5\pm 3.5\text{~GeV}$, where the uncertainty is
mainly due to the error in the top quark mass.

\section{Final remarks and conclusions}

We have argued that the upper bound on the lightest Higgs boson
mass in the NMSSM attains its maximal value in the strong Yukawa
coupling limit. In the considered limit all solutions of
renormalization group equations gathered near the quasi--fixed
points. If the scale of supersymmetry breaking is much larger than
the electroweak one the perturbation theory can be used for the
calculation of the Higgs boson masses. However even when $M_S\gg
M_Z$ the lightest CP--even Higgs boson mass in the NMSSM is
appreciably smaller than its upper bound in the dominant part of
the parameter space. Besides in the strong Yukawa coupling limit
within the NMSSM with a minimal set of fundamental parameters the
self--consistent solution does not exist. Moreover the $Z_3$
symmetry of the NMSSM superpotential leads to the domain wall
problem.

We have suggested such a modification of the NMSSM that allows to
get the self--consistent solution in the strong Yukawa coupling
limit and at the same time to avoid the domain wall problem. The
superpotential of the modified NMSSM (MNSSM) includes the bilinear
terms which break the $Z_3$ symmetry. We have studied the spectrum
of the Higgs bosons in the MNSSM. The qualitative pattern of the
particle spectrum is most sensitive to the choice of two
parameters -- $\mu'$ and $M_S$. The limit $\mu'\gg M_S$, when the
CP--even and CP--odd scalar fields become very heavy, corresponds
to the minimal SUSY model. In the most interesting region of the
parameter space, where the mass of the lightest Higgs boson is
larger than that in the MSSM, the Higgs boson mass matrix has the
hierarchical structure and can be diagonalised using the method of
perturbation theory. The heaviest particle in this region of the
MNSSM parameters is the CP--even Higgs boson corresponding to the
neutral scalar field $Y$ and the heaviest fermion is $\tilde{Y}$
which is the superpartner of the singlet field $Y$. The lightest
Higgs boson mass in the considered model may reach
$127\text{~GeV}$ even for the comparatively low value of
$\tan\beta\simeq 1.9$ and does not exceed $130.5\pm
3.5\text{~GeV}$.

The obtained upper bound on the mass of the lightest CP--even
Higgs boson is not an absolute one in the supersymmetric models.
For instance, the upper bound on $m_h$ increases if new
$5+\bar{5}$ supermultiplets appear in the NMSSM at the SUSY
breaking scale. These multiplets change the evolution of gauge
couplings. Their values at the intermediate scale rise if a number
of new supermultiplets increases. For this reason the allowed
region of the Yukawa couplings at the electroweak scale is
expanded. It leads to the growth of the upper bound on $m_h$ as
the number of $5+\bar{5}$ supermultiplets rises (see Fig. 4). The
investigations performed in \cite{36} showed that the introduction
of four or five $5+\bar{5}$ supermultiplets raises the theoretical
bound on the lightest Higgs boson mass up to $155\text{~GeV}$. If
more than five multiplets are introduced at the SUSY breaking
scale then the gauge couplings blow up before the Grand
Unification scale and perturbation theory is not valid at $q^2\sim
M_X^2$.

Recently the upper bound on the lightest Higgs boson mass in more
complicated SUSY models has been analysed
[\citen{44}--\citen{46}]. In particular, in addition to the gauge
singlet superfield three $SU(2)$ triplets $\hat{T}_i$ with
different hypercharges can be introduced into the Higgs boson
superpotential:
\begin{equation}
W_{\text{Higgs}}=\lambda\hat{Y}(\hat{H}_1\hat{H}_2)+\lambda_1(\hat{H}_1\hat{T}_0\hat{H}_2)+
\chi_1(\hat{H}_1\hat{T}_1\hat{H}_1)+\chi_2(\hat{H}_2\hat{T}_{1}\hat{H}_2)+\dots\,.\label{25}
\end{equation}
As a result the expression for the upper bound on $m_h$ changes
\begin{equation}
m_h^2\le M_Z^2\cos^2
2\beta+\Biggl[\left(\frac{\lambda^2}{2}+\frac{\lambda_1^2}{4}\right)
\sin^2 2\beta+ 2\chi_1^2\cos^4\beta+2\chi_2^2\sin^4\beta\Biggr]
v^2+\Delta. \label{26}
\end{equation}
The appearance of triplet superfields destroys the gauge coupling
constant unification at high energies. In order to restore the
unification scheme of the electroweak and strong interactions one
has to add several $SU(3)$ multiplets, for example four
$(3+\bar{3})$, which do not participate in the $SU(2)\otimes U(1)$
interactions. A numerical analysis [\citen{45},\citen{451}]
reveals that the unification of gauge couplings then occurs at the
scale $\tilde{M}_X\sim 10^{17}\text{~GeV}$. As one can see from
Fig. 5 (see also \cite{45}) the upper bound on the lightest Higgs
boson mass rises with growth of $\tan\beta$ and for $\tan\beta\gg
1$ can be approximately equal to $190\text{~GeV}$ \cite{451}.

Also the upper bound on $m_h$ is raised if in the MSSM a fourth
generation of the quarks and leptons exists \cite{46}. However up
to now there has not been found any evidence of the existence of
the fourth generation in the SM or the MSSM. Moreover, new
particles give considerable contributions to the electroweak
observables which upset the agreement between theoretical
predictions and the results of experimental measurements. Thus the
growth of the upper bound on the lightest Higgs boson mass in the
supersymmetric models is usually accompanied by substantial
increase in the number of particles that may be considered as the
main drawback of these models.

\section*{Acknowledgements}

The authors are grateful to D.I.Kazakov, L.B.Okun, and
M.I.Vysotsky for stimulating discussions. One of the authors
(R.B.N.) thanks the Italian National Institute of Nuclear Physics
(Ferrara Division), where a considerable part of the
investigations was performed, for their hospitality. Our work was
supported by the Russian Foundation for Basic Research (RFBR)
(projects 00-15-96786 and 00-15-96562).

\newpage

{\bfseries Table.} Spectrum of the superparticles and Higgs bosons
for $m_t(174\text{~GeV})=165\text{~GeV}$, $A=m_0=0$ and for
different initial values of $h_t^2(0)$ and $\lambda^2(0)$ (all
mass parameters are given in GeVs).

\begin{center}
\begin{tabular}{|c|c|c|c|c|c|c|}
\hline
&\multicolumn{3}{|c|}{$\mu_{\text{eff}}<0$}&\multicolumn{3}{|c|}{$\mu_{\text{eff}}>0$}\\
\hline $\lambda^2(0)$&2&10&10&2&10&10\\ \hline
$h_t^2(0)$&10&10&2&10&10&2\\ \hline
$M_{1/2}$&-392.8&-392.8&-392.8&-392.8&-392.8&-392.8\\ \hline
$\tan\beta$&1.736&1.883&2.439&1.736&1.883&2.439\\ \hline
$\mu_{eff}$&-771.4&-727.8&-641.8&772.4&728.6&642.3\\ \hline
$B_0$&622.5&1008.0&886.2&-988.1&-1629.1&-1583.3\\ \hline
$y$&-0.0014&-0.0015&-0.0012&-0.0003&-0.0004&-0.0005\\ \hline
$\mu'(t_0$)&1693.9&1671.5&1749.8&-1941.4&-1899.8&-1943.1\\ \hline
${\bf m_h (t_0)}$&{\bf 123.6}&{\bf 134.1}&{\bf 137.6} &{\bf
112.4}&{\bf 125.0}&{\bf 131.2}\\ {\bf (one--loop)}&&&&&&\\ \hline
${\bf m_h (t_0)}$&{\bf 113.0}&{\bf 124.4}&{\bf 127.8} &{\bf
105.5}&{\bf 118.4}&{\bf 123.6}\\ {\bf (two--loop)}&&&&&&\\ \hline
$M_3(1$\,TeV)&1000&1000&1000&1000&1000&1000\\ \hline
$m_{\tilde{t}_1}(1$\,TeV)&891.6&890.2&890.5&837.0&840.6&853.5\\
\hline
$m_{\tilde{t}_2}(1$\,TeV)&622.2&630.3&648.5&693.8&695.1&696.4\\
\hline $m_H(1$\,TeV)&961.0&896.2&758.5&963.3&898.5&761.1\\ \hline
$m_S(1$\,TeV)&1999.8&2147.4&2187.2&2405.3&2623.4&2663.8\\ \hline
$m_{A_1}(1$\,TeV)&1374.8&1123.2&1294.0&1390.6&953.9&965.1\\ \hline
$m_{A_2}(1$\,TeV)&949.8&857.6&735.6&951.6&704.3&674.3\\ \hline
$m_{\tilde{\chi}_1}(t_0)$&160.1&160.0&159.9&164.6&164.6&164.4\\
\hline
$m_{\tilde{\chi}_2}(t_0)$&311.9&311.1&309.4&328.1&327.8&326.4\\
\hline
$|m_{\tilde{\chi}_3}(1$\,TeV)$|$&795.8&753.7&665.8&797.2&755.1&668.1\\
\hline
$|m_{\tilde{\chi}_4}(1$\,TeV)$|$&807.8&764.7&677.1&800.9&755.9&666.7\\
\hline
$|m_{\tilde{\chi}_5}(1$\,TeV)$|$&1711.2&1700.7&1790.0&1960.7&1931.8&1986.5\\
\hline
$m_{\tilde{\chi}^{\pm}_1}(t_0)$&311.6&310.7&309.0&328.1&327.8&326.4\\
\hline
$m_{\tilde{\chi}^{\pm}_2}(1$\,TeV)&806.0&763.3&676.7&800.4&757.0&669.0\\
\hline
\end{tabular}
\end{center}

\newpage
{\bf \Large Figure captions}

\noindent {\bf Figure 1.} The particle spectrum (in GeVs) in the
modified NMSSM as a function of $z=\mu'/1\text{\,TeV}$ and
$x=A/M_{1/2}$ for $h^2_t(0)=\lambda^2(0)=10$, $m^2_0=0$,
$M_3=1\text{\,TeV}$, and $\mu_{\text{eff}}<0$. Thick and thin
curves in Fig.1{\itshape a} and 1{\itshape b} correspond to the
lightest Higgs boson mass calculated in the one-- and two--loop
approximation, respectively. In Fig.1{\itshape c} and 1{\itshape
d} thick and thin curves reproduce the dependence of CP--even
Higgs boson masses $m_S$ and $m_H$ on $z$ and $x$ while dotted and
dashed curves represent the CP--odd Higgs boson masses $m_{A_1}$
and $m_{A_2}$ as a functions of these parameters. The
dashed--dotted curve in Fig.1{\itshape c} corresponds to the mass
of the heaviest neutralino.\\

\noindent {\bf Figure 2.} The particle spectrum (in GeVs) in the
modified NMSSM as a function of $z=\mu'/1\text{\,TeV}$ and
$x=A/M_{1/2}$ for $h^2_t(0)=\lambda^2(0)=10$, $m^2_0=0$,
$M_3=1\text{\,TeV}$, and $\mu_{\text{eff}}>0$. The notations are
the same as in Fig.1.\\

\noindent {\bf Figure 3.} Upper bound on the lightest Higgs boson
mass (in GeVs) in the MSSM (lower curve) and in the modified NMSSM
(upper curve) as a function of $\tan\beta$ for
$M_3=2\text{~TeV}$.\\

\noindent {\bf Figure 4.} The dependence of the upper bound on the
lightest Higgs boson mass (in GeVs) in the NMSSM (solid and
long--dashed curves) and in the MSSM (short--dashed curve) on the
value of $\tan\beta$. The solid curve is the upper bound on $m_h$
in the NMSSM with four additional $5+\bar{5}$ multiplets while the
long--dashed curve is that one in the ordinary NMSSM (see
\cite{36}).\\

\noindent {\bf Figure 5.} Theoretical upper bound on the lightest
Higgs boson mass in the supersymmetric model, which contains
singlet and three $SU(2)$ triplet fields in the Higgs sector and
four $3+\bar{3}$ colour multiplets, as a function of $\tan\beta$
for $\lambda_1=0$ (see \cite{45}).

\newpage

\vspace*{-10mm}\noindent
\includegraphics[width=159mm, keepaspectratio=true]{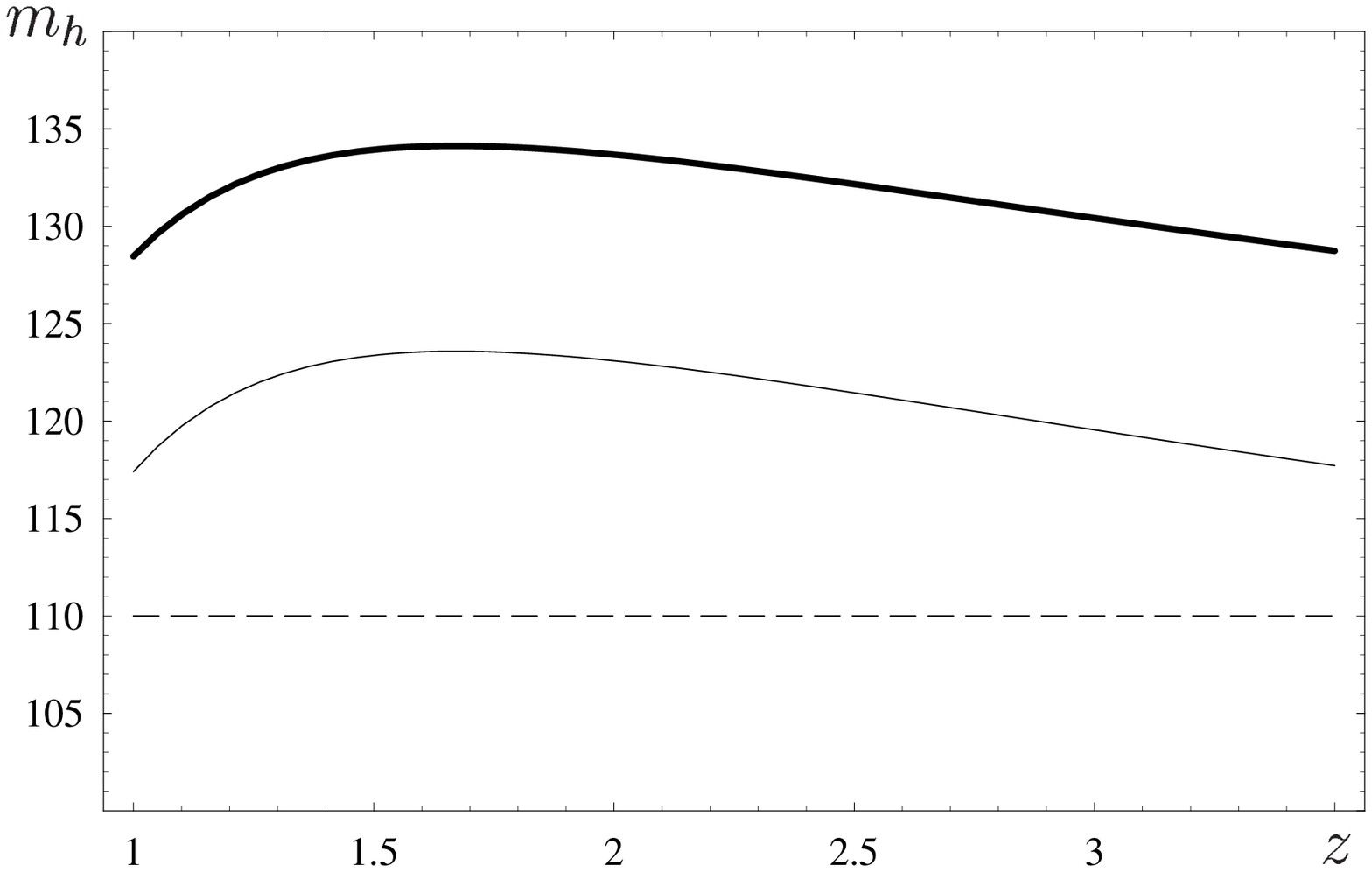}

\vspace{5mm}\hspace*{70mm}{\large\bfseries Fig.1a.}

\vspace{20mm}

\noindent
\includegraphics[width=159mm, keepaspectratio=true]{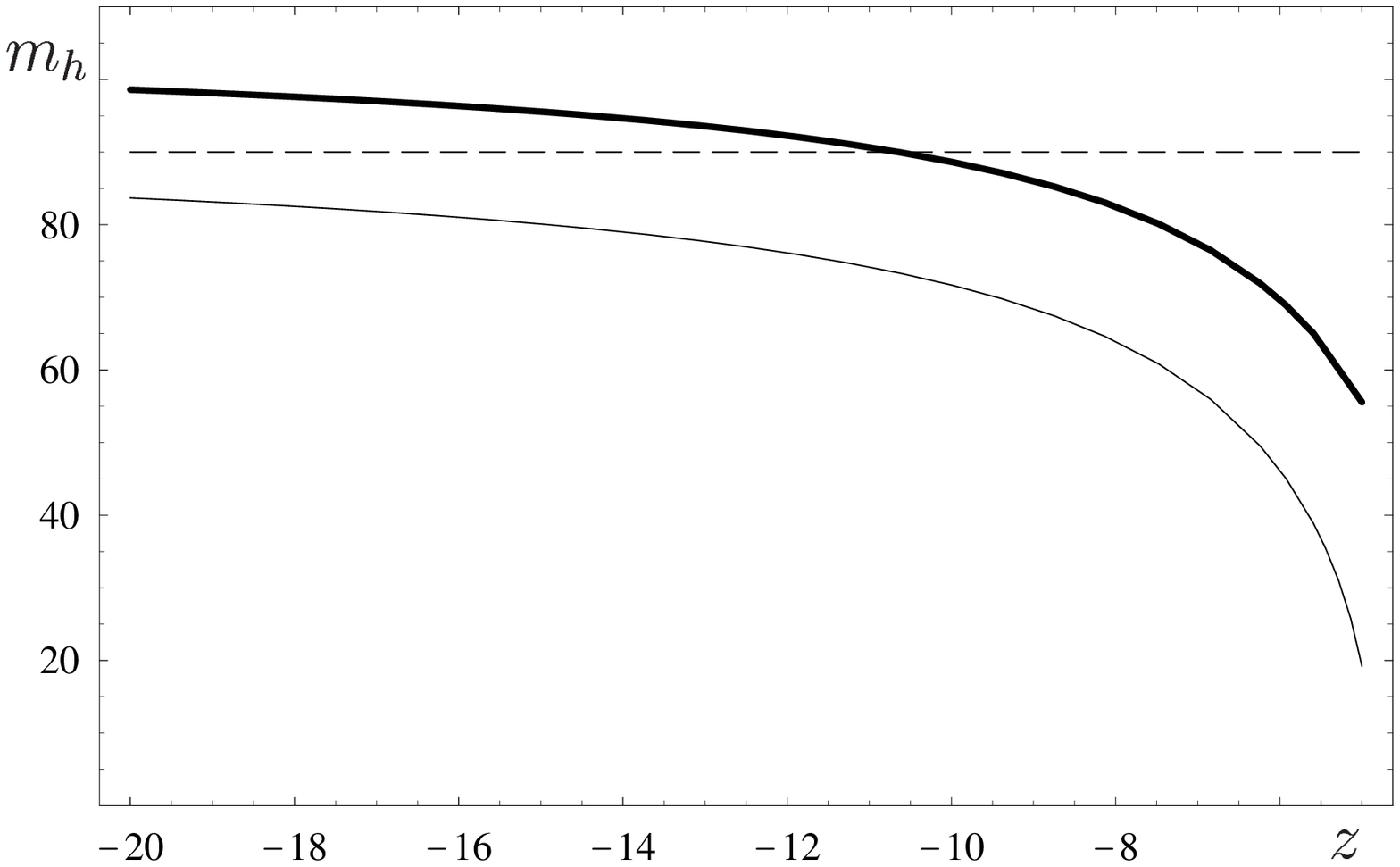}

\vspace{5mm}\hspace*{70mm}{\large\bfseries Fig.1b.}

\newpage

\noindent
\includegraphics[width=159mm, keepaspectratio=true]{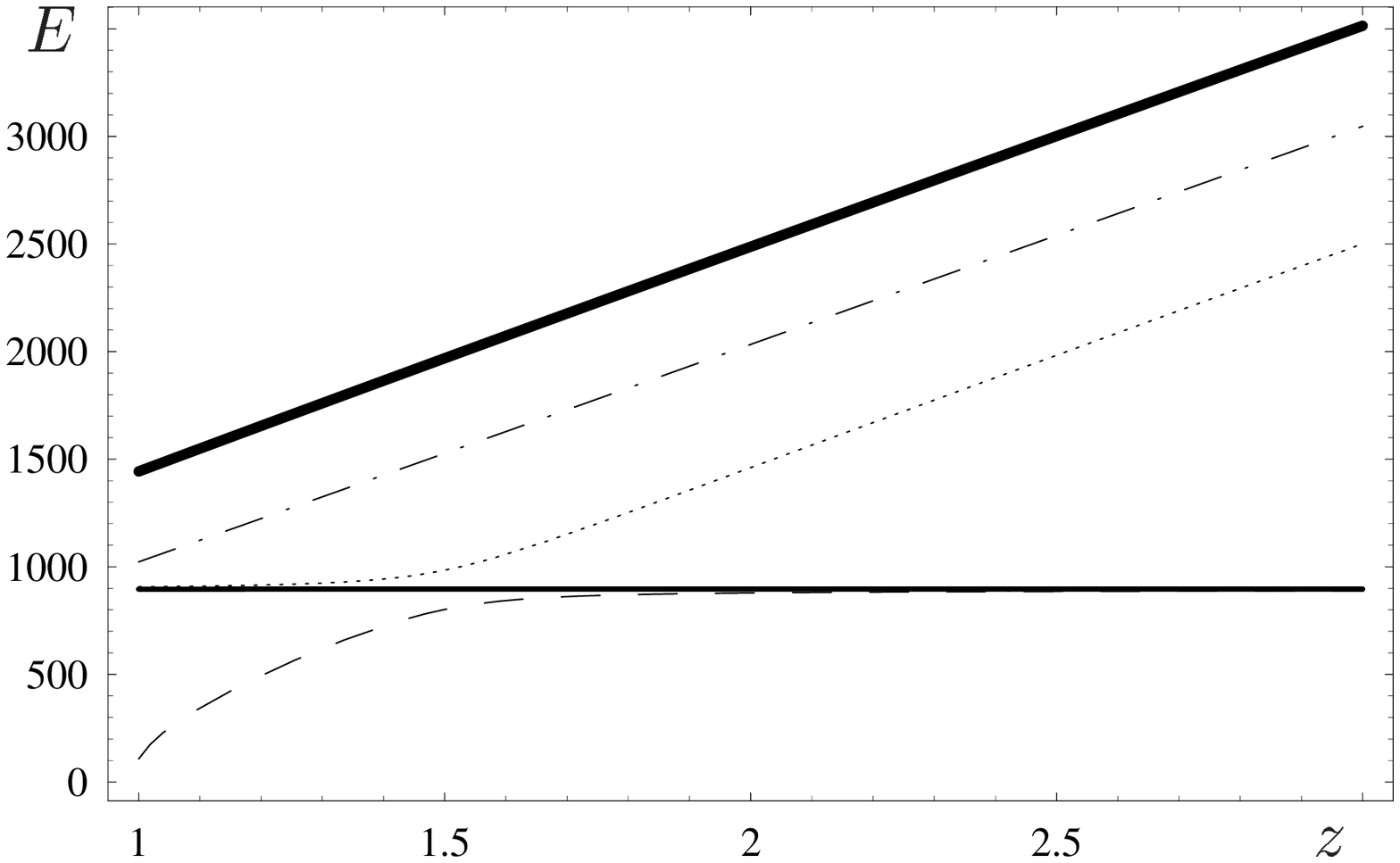}

\vspace{5mm}\hspace*{70mm}{\large\bfseries Fig.1c.}

\vspace{20mm}

\noindent
\includegraphics[width=159mm, keepaspectratio=true]{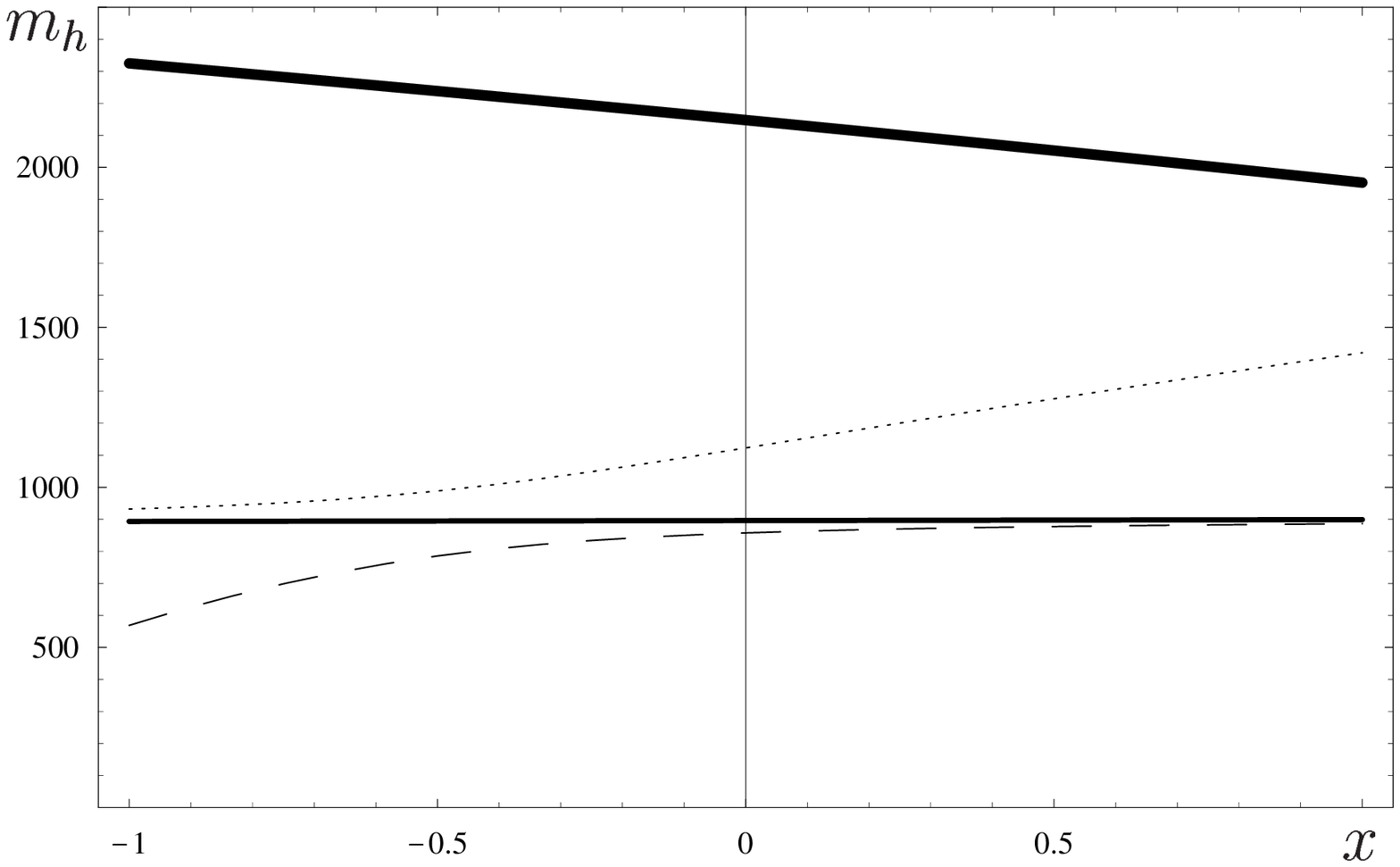}

\vspace{5mm}\hspace*{70mm}{\large\bfseries Fig.1d.}

\newpage

\vspace*{-10mm}\noindent
\includegraphics[width=159mm, keepaspectratio=true]{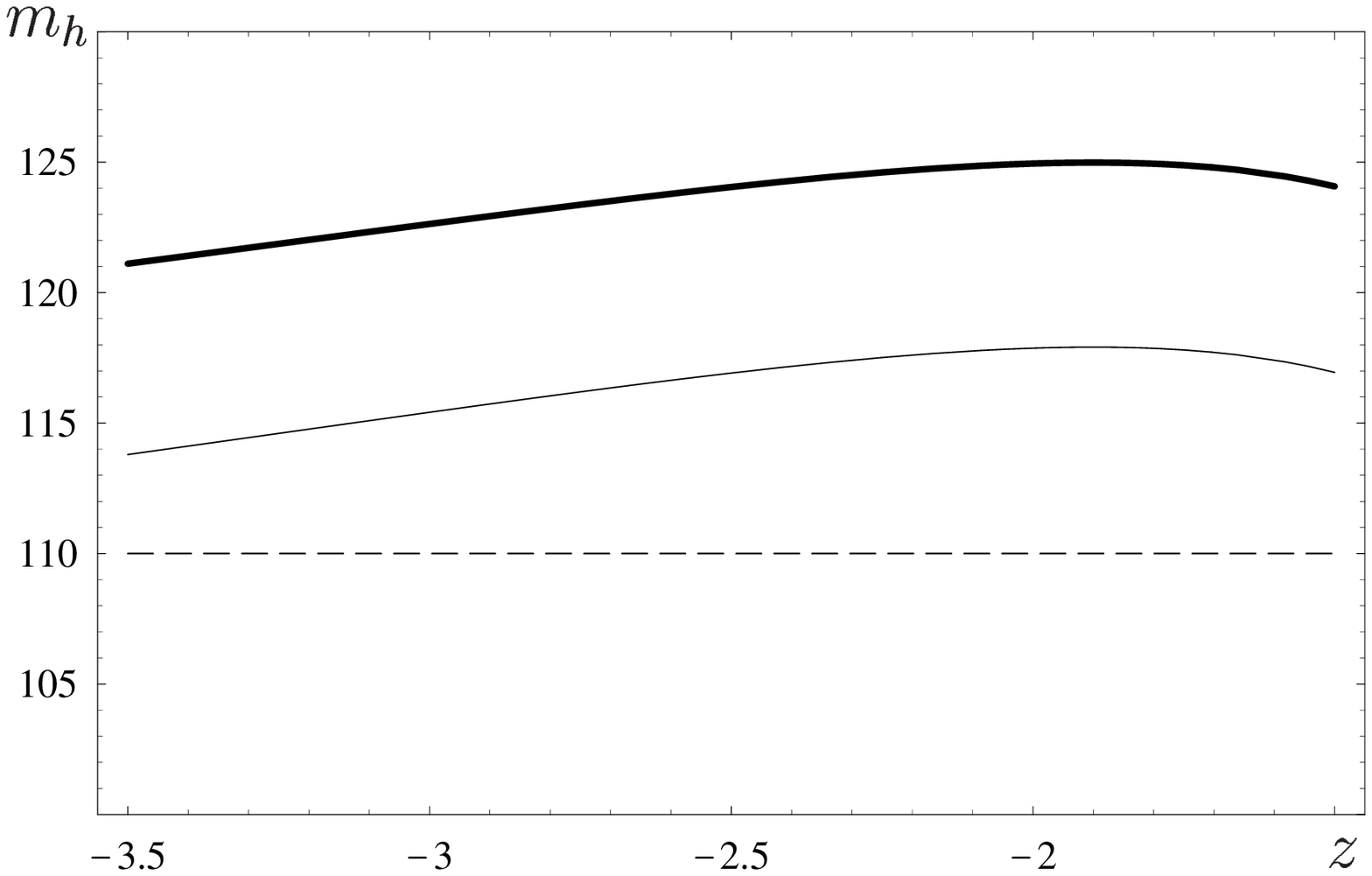}

\vspace{5mm}\hspace*{70mm}{\large\bfseries Fig.2a.}

\vspace{15mm}

\noindent
\includegraphics[width=159mm, keepaspectratio=true]{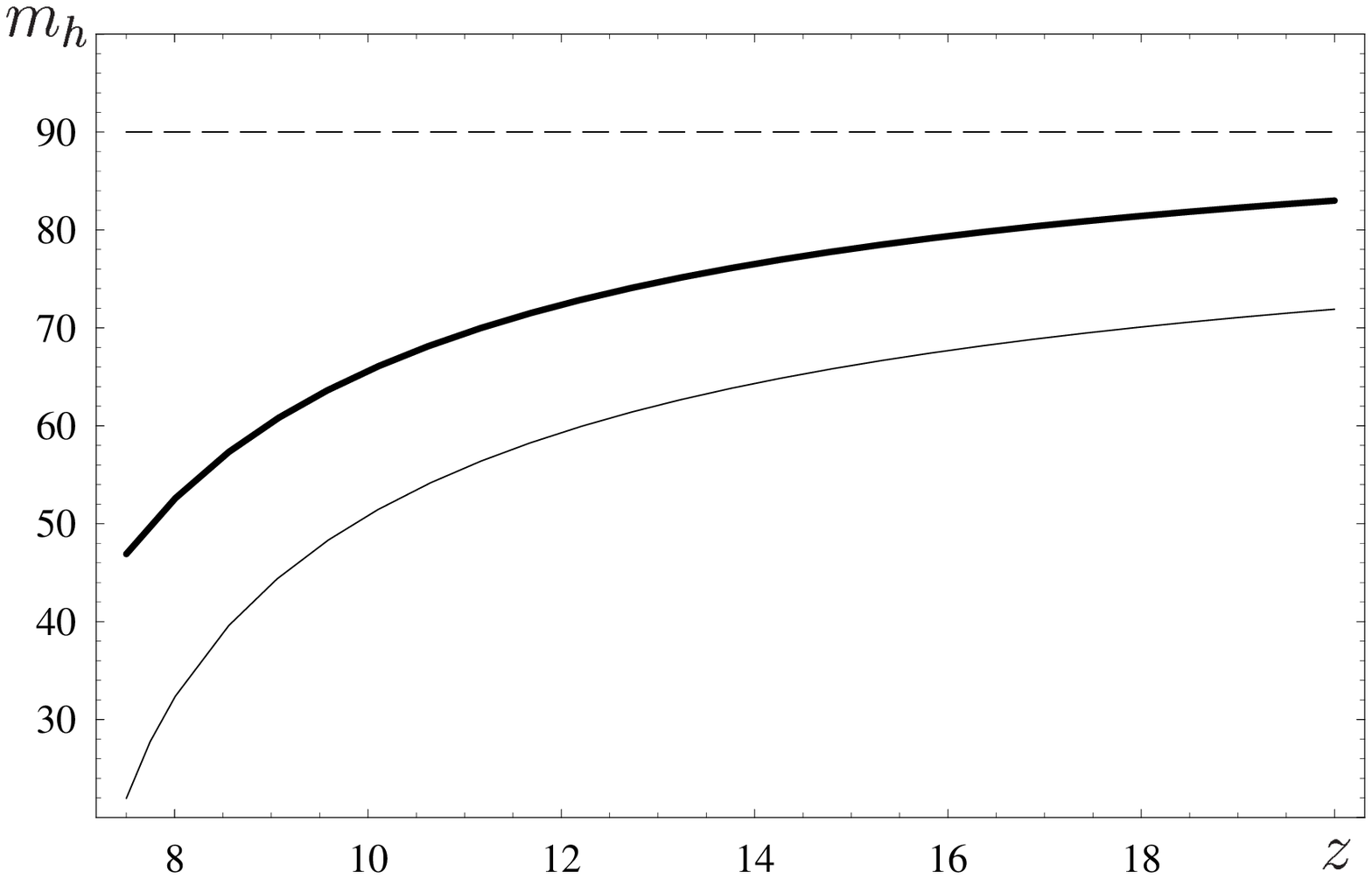}

\vspace{5mm}\hspace*{70mm}{\large\bfseries Fig.2b.}

\newpage

\vspace*{-10mm}\noindent
\includegraphics[width=159mm, keepaspectratio=true]{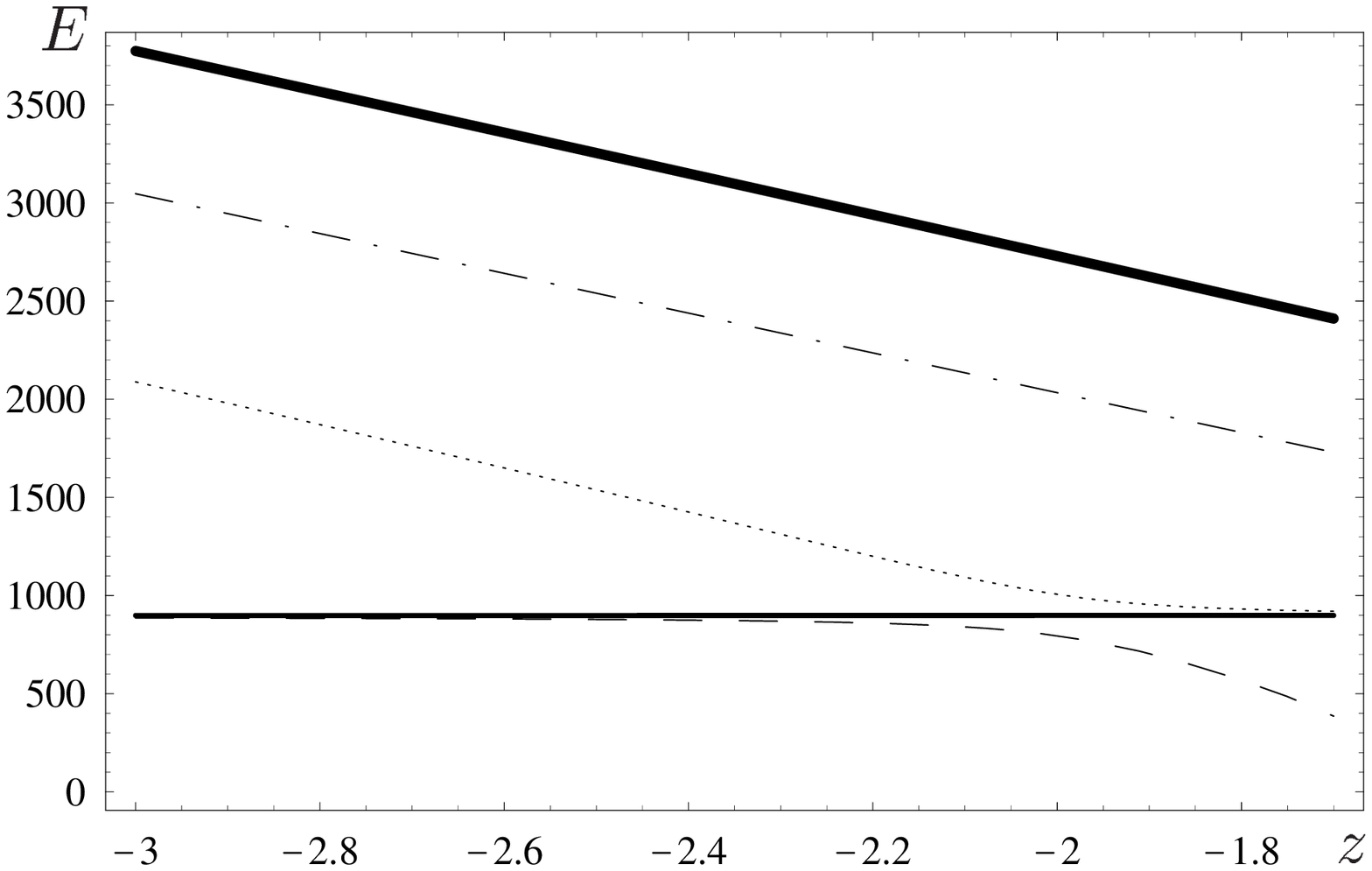}

\vspace{5mm}\hspace*{70mm}{\large\bfseries Fig.2c.}

\vspace{15mm}

\noindent
\includegraphics[width=159mm, keepaspectratio=true]{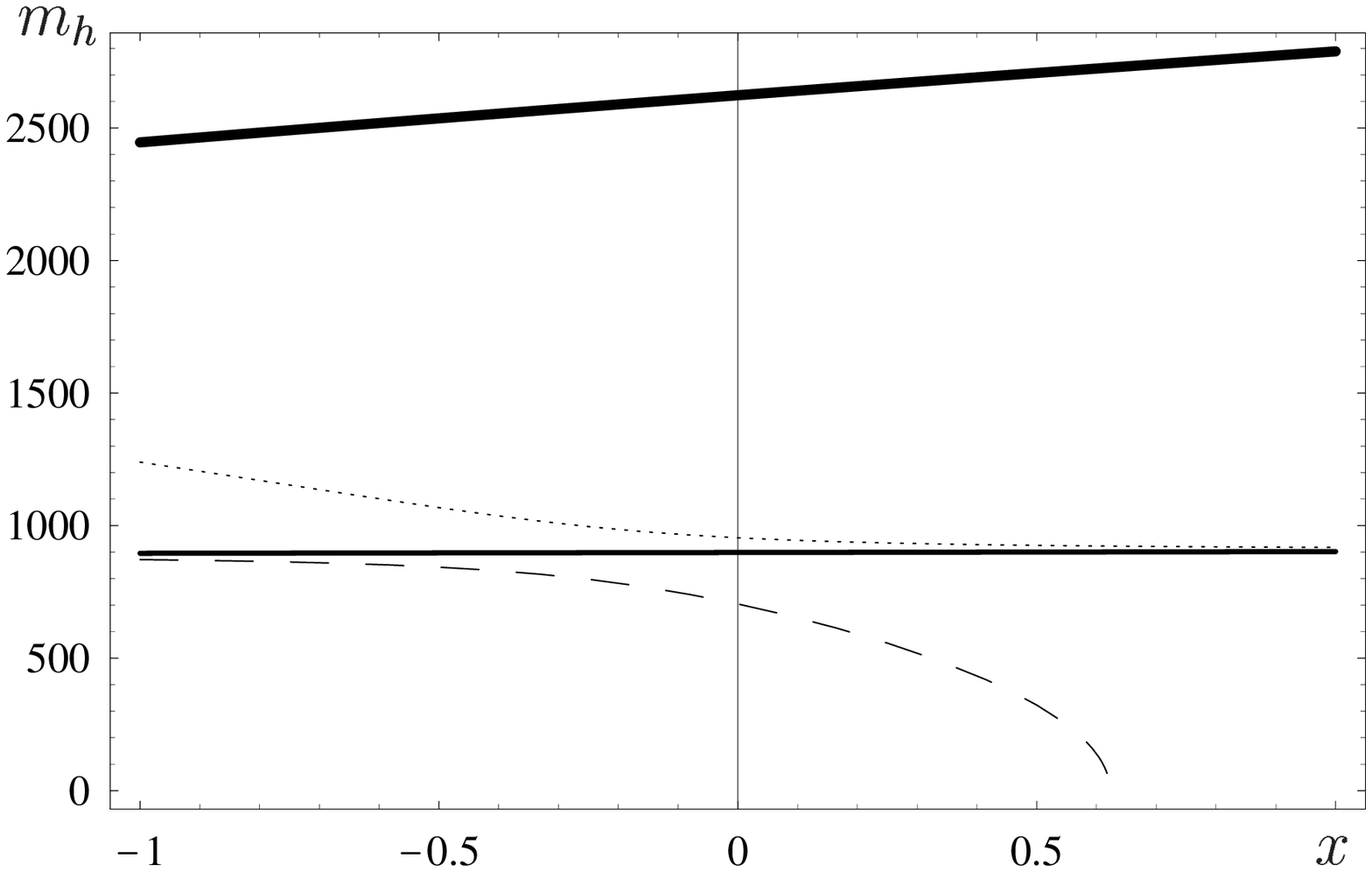}

\vspace{5mm}\hspace*{70mm}{\large\bfseries Fig.2d.}

\begin{landscape}

\vspace*{-9mm} \noindent\hspace*{3mm}
\includegraphics[width=145mm, keepaspectratio=true, angle=-90]{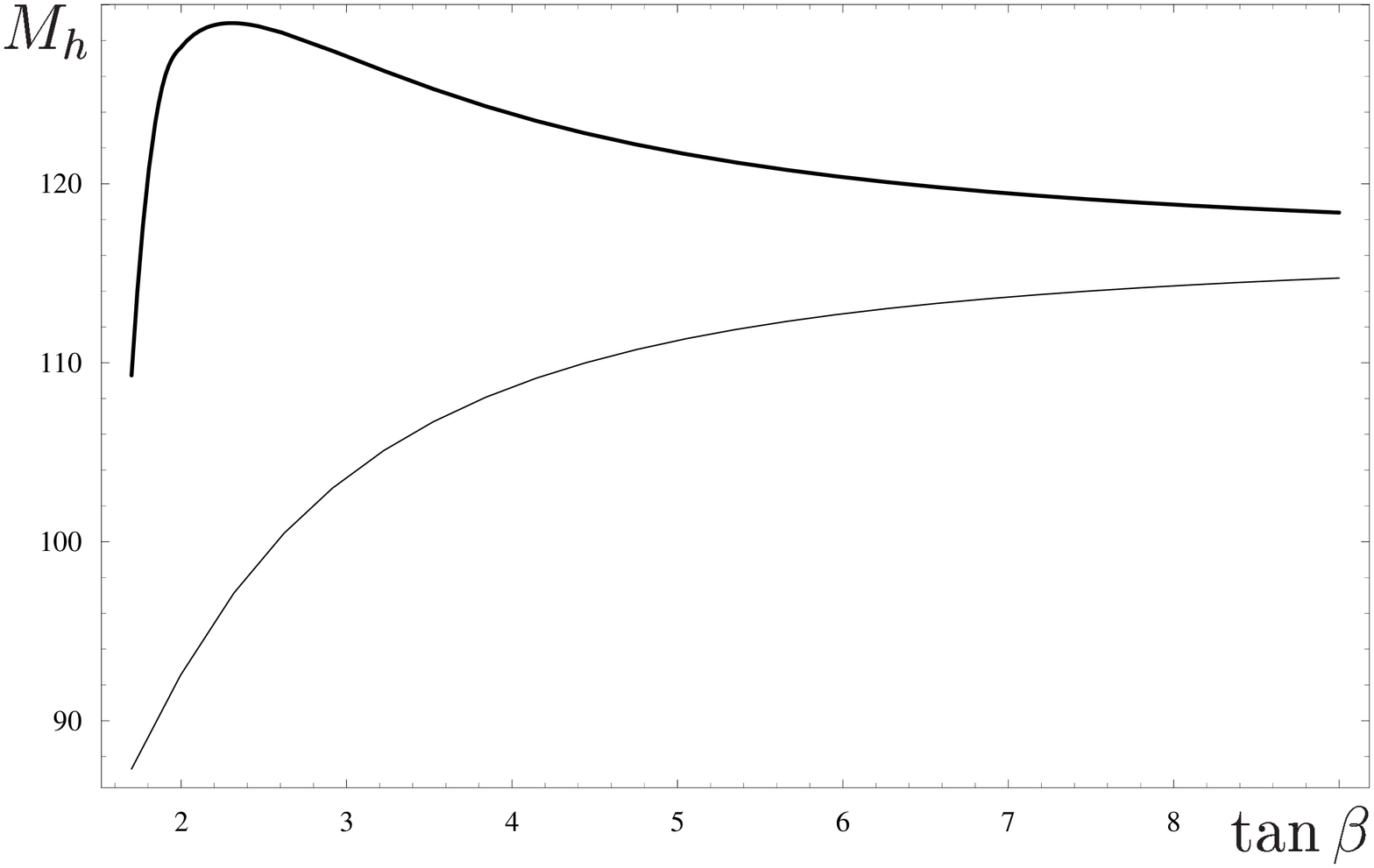}

\vspace{4mm}\hspace*{117.5mm}{\large\bfseries Fig.3.}

\end{landscape}

\noindent\includegraphics{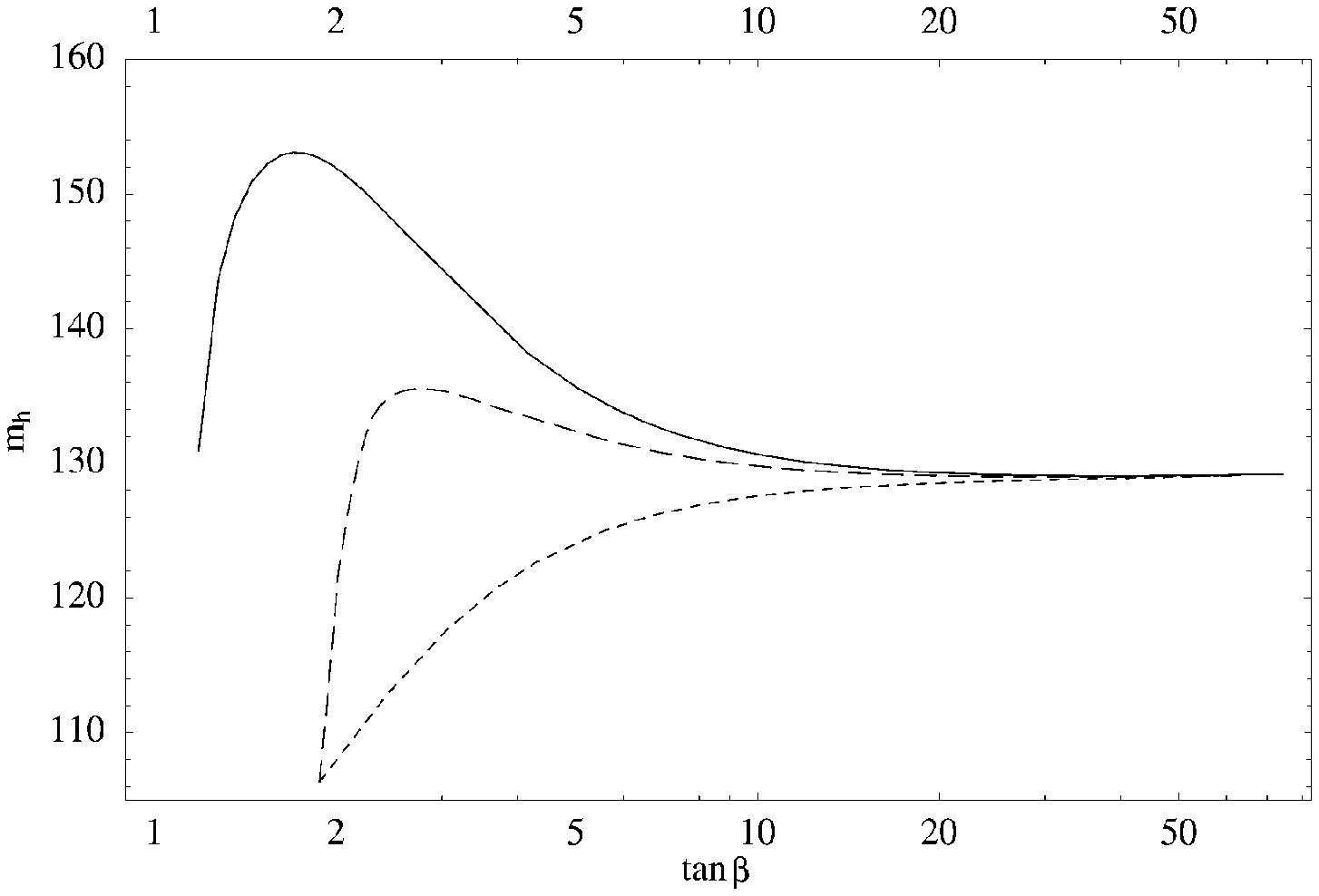}

\vspace{5mm}\hspace*{70mm}{\large\bfseries Fig.4.}

\vspace*{20mm}

\noindent\hspace*{27mm}\includegraphics{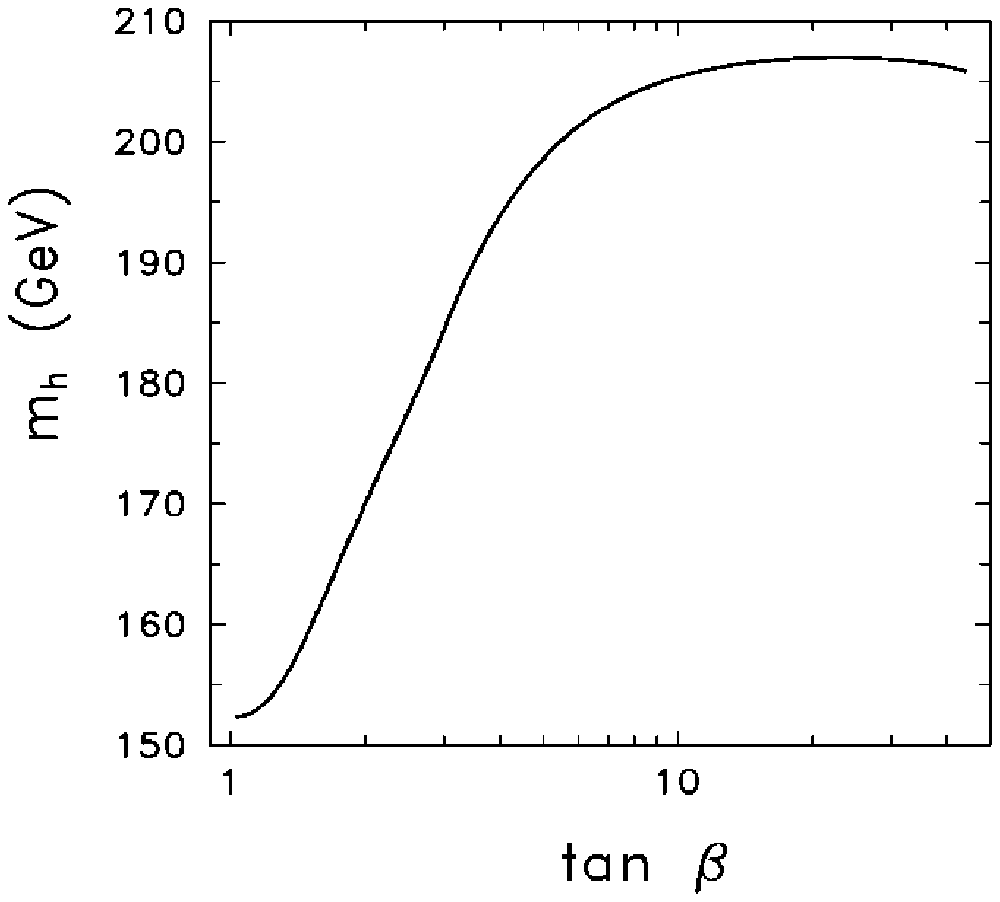}

\vspace{5mm}\hspace*{70mm}{\large\bfseries Fig.5.}

\end{document}